%% file: paper_v2.tex
\documentclass[a4paper,11pt]{article}
\pdfoutput=1

\usepackage{a4wide}
\usepackage{amsfonts}
\usepackage{amssymb}
\usepackage{amsmath}
\usepackage{braket}
\usepackage{slashed}
\usepackage{float}
\usepackage{graphicx}
\usepackage{tikz}
\usetikzlibrary{patterns,arrows}
\usepackage[small,bf]{caption}
\usepackage{booktabs}
\usepackage{hyperref}
\usepackage{color}
\usepackage{cite} 
\usepackage{ulem}

\allowdisplaybreaks[2]
\numberwithin{equation}{section}

\DeclareMathOperator{\diff}{\text{d}}

\begin{document}
\begin{titlepage}

\begin{center}
{
\bf\LARGE Detection prospects for the Cosmic Neutrino Background using matter interferometers
}
\\[8mm]
Chrisna Setyo Nugroho$^{\, a, \, b}$ \footnote{E-mail: \texttt{setyo13nugros@gmail.com, chrisna@apps.ipb.ac.id}}
and 
Martin Spinrath$^{\, c}$ \footnote{E-mail: \texttt{spinrath@phys.nthu.edu.tw}}
\\[1mm]
\end{center}

\vspace*{0.50cm}
{\centering \it
$^{a}$ Theoretical Physics Division, Department of Physics, \\
	IPB University (Bogor Agricultural University), \\
	Jl. Meranti, Kampus IPB Dramaga, Bogor 16680, Indonesia\\[0.2cm] 
$^{b}$ Department of Physics, National Taiwan Normal University, Taipei 116, Taiwan   \\[0.2cm] 
$^{c}$ Department of Physics, National Tsing Hua University,\\ Hsinchu, 30013, Taiwan\\
}
\vspace*{1.20cm}

\begin{abstract}
\noindent
In this paper we discuss how the Cosmic Neutrino Background can affect the measured phase
difference in a matter interferometer.
This phase is proportional to a difference in potential energies along the two
interferometer paths. The relevant potentials here are the well-known neutrino matter
potential and a potential related to the Stodolsky effect. We show how they
can be rewritten in terms of scalar potentials, pseudo magnetic fields and
spin-spin interactions.
Unfortunately, current technology is unlikely to detect this effect and we discuss
prospects for the future. We also briefly comment on fermionic Dark Matter
which can give rise to very similar effects which can easily be larger than the neutrino case. 
\end{abstract}

\end{titlepage}
\setcounter{footnote}{0}

\section{Introduction}

The standard big bang model of cosmology predicts that in the early universe neutrinos decoupled
from the hot plasma and that they should still be around today. They form the so-called
Cosmic Neutrino Background (CNB) analogous to the Cosmic
Microwave Background. In the standard scenario we expect these neutrinos to have a
temperature today of about $T_\nu \simeq 1.95 \text{ K} \simeq 1.68 \times 10^{-4} \text{ eV}$.
The average number density per flavor and helicity is $n_0 \simeq 56 \text{ cm}^{-3}$. But this number
can be enhanced locally by a factor up to $\mathcal{O}(10)$, for some recent studies,
see, for instance, \cite{Zimmer:2024max,Holm:2024zpr,Worku:2024kwv}.
Due to CP violation in the Standard Model of particle physics (SM) one would expect
a very small neutrino-antineutrino asymmetry but
in SM extensions that could be enhanced and the 
neutrino lepton number flavor asymmetries in the Universe today can be as large
as $\sim 35 \text{ cm}^{-3}$~\cite{Domcke:2025jiy}.

It would be very interesting and insightful for particle physics and cosmology if one could
probe the properties of the CNB in a laboratory. There have been many proposals, for
(non-exhaustive lists of) references,
see, for instance, \cite{Bauer:2022lri,Shergold:2023fdd,delCastillo:2025qnr}. Also the authors
of this paper have discussed possibilities in that direction, see
\cite{Domcke:2017aqj,Asteriadis:2022zmo,Nugroho:2023cun,Chen:2025tlx}.
In this paper we want to study the prospects to utilize matter interferometers
for the detection of the CNB. The physical origins of the effect discussed here can be traced back to
a potential similar to the matter potential which results in the
Mikheev--Smirnov--Wolfenstein (MSW) effect~\cite{Wolfenstein:1977ue,Mikheev:1986wj}
and a potential originating from the Stodolsky effect \cite{Stodolsky:1974aq}.
For the sake of simplicity, in the following we will sometimes just call it MSW or Stodolsky effect.
Recently, it was pointed out that due to scattering effects near the Earth surface,
asymmetries in the CNB number densities of the order of $10^{-8}$ can be generated 
\cite{Arvanitaki:2022oby,Huang:2024tog,Gruzinov:2024ciz,Kalia:2024xeq} which affect the potentials
and will be crucial
to create an observable effect in an interferometer, in particular, for the neutrino matter potential case. 

Furthermore,
Dark Matter (DM) can have an analogous effect to the Stodolsky effect, the Dark Stodolsky 
effect~\cite{Rostagni:2023eic}. And as we will see there is also an analogue to the neutrino matter potential
for the MSW effect which could lead to a Dark MSW effect. We will briefly comment on these but
the main focus of this work is on the CNB.

Optical interferometry is an old technique utilized in high precision experiments. It relies on the
measurement of the phase difference of light traversed along two different, often perpendicular arms of the
interferometer. This technology has experienced tremendous
improvements in recent decades which allowed physicists to build ever more advanced interferometers and
one of its most famed recent successes was the discovery of gravitational waves by the LIGO/VIRGO
collaboration \cite{LIGOScientific:2016aoc}. But it does not end there and the community is already
planning so-called
third generation laser interferometers such as the Einstein Telescope \cite{Punturo:2010zz} and Cosmic
Explorer \cite{Reitze:2019iox}. 

Here we want to focus on a different kind of interferometer, matter interferometers.
Recent developments in quantum technology such as quantum optics
and quantum metrology have enabled physicists to develop matter wave interferometers such as neutron
interferometers as well as atom interferometers which demonstrate an unprecedented sensitivity to
observe many important phenomena such as the Aharonov--Casher effect of neutral particles
\cite{Aharonov:1984xb} and the gravitational Aharonov--Bohm effect \cite{Overstreet:2021hea}.   
For that reason atom interferometers have invited a growing interest from the gravitational wave 
community. Several 
proposals such as AION \cite{Badurina:2019hst}, AICE \cite{Baynham:2025pzm}, MAGIS \cite{Graham:2017pmn,Coleman:2018ozp},
ELGAR \cite{Canuel:2019abg}, MIGA \cite{Canuel:2017rrp}, and ZAIGA \cite{Zhan:2019quq} have been made
and there is a flurry of developments going on. 
The main purpose of those is usually to
detect gravitational waves from different sources but also other applications like searches for
ultralight, wave-like DM have been discussed.
In this paper we discuss a different potential application, namely to look for the
CNB using matter interferometers. But the physics we discuss here
can also apply to certain types of DM as we will mention briefly as well.
In \cite{Pinheiro:2025kkr} they consider neutrinos acting on matter interferometers but they focus on
 decoherence which we do not discuss here.

The rest of the paper is organized as follows: In Sec.~\ref{sec:Matter_Interferometers}
we provide a brief description on how a matter interferometer works and provide
the crucial formula which relates an observable phase difference to the potentials
experienced by the matter along the interferometer arms.
In Sec.~\ref{sec:Potentials} we discuss then the effect of neutrinos on the matter
in the interferometer arms which can be described using potentials.
We will then estimate the effect size of these potentials in Sec.~\ref{sec:Results} 
and compare it to current experimental sensitivities before we summarize
and conclude in Sec.~\ref{sec:Summary}.

\section{Matter interferometers}
\label{sec:Matter_Interferometers}

In principle, a matter interferometer exploits the wave nature of any fundamental building block of matter.
To simplify the description we will use atoms here as example, but one could equally well replace
them by electrons, neutrons or even molecules. The fundamental principles remain the same.

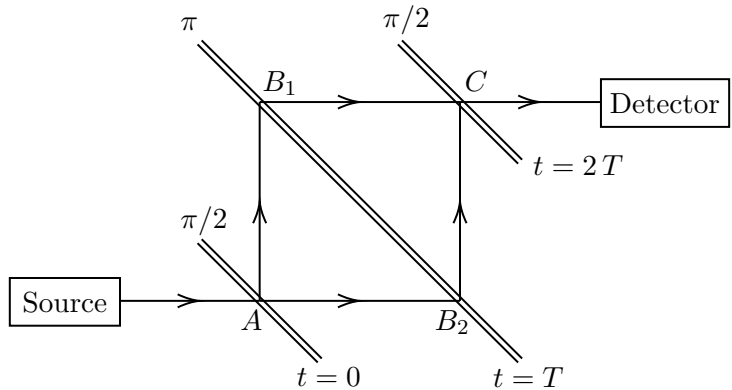
\begin{figure}
    \centering
    \input{figs/Setup.tikz}
    \caption{Simplified sketch of a matter interferometer. At the points $A$ and $C$, that is
    at $t=0$ and $t = 2 \, T$ a $\pi/2$ pulse is applied which acts as a beam splitter or merger.
    At $B_1$ and $B_2$ a $\pi$ pulse is applied which leads to an inversion of the ground and excited
    state.
    }
    \label{Fig:Setup}
\end{figure}

Atom interferometers probe the phase shift carried by the atoms during their propagation through the
two interferometer
arms. In its simplest realization, the atoms are modeled as a two level quantum system.
There is the ground state $\ket{g}$ and an excited state $\ket{e}$. Initially, an atom is prepared in
the ground state $\ket{g}$. Next, at $t = 0$, a so-called $\pi/2$ laser pulse is directed towards the
atom which
makes both states $\ket{g}$ and $\ket{e}$ equally populated. The wave function of the atom now is
in a coherent superposition between $\ket{g}$ and $\ket{e}$. Thus, the $\pi/2$ pulse acts as a beam
splitter that divides the atoms into two equal matter-waves which can be spatially separated.
Subsequently, at $t = T$, a second laser
pulse called $\pi$-type is applied towards both matter-waves which switch the internal state $\ket{g}$ to $\ket{e}$ and vice versa. Finally, after the matter-waves are reflected by the ``mirror'' $\pi$ pulse, at
$t = 2 \, T$, the matter-waves are recombined using another $\pi/2$ light pulse. At the output port of this
matter-wave interferometer, the phase shift picked up by the atom during its journey is measured.
We show a very rough sketch in Fig.~\ref{Fig:Setup}.

The phase shift between the two arms can be expressed in terms of a path integral
over the two interferometer arms, see, for instance, \cite[Chapter 1.2]{Rauch:2015jkh},
\begin{align}
	\Delta \Phi  = - \frac{1}{\hbar} \oint V(\vec{x}(t)) \diff t \;,
\end{align}
which works for time-independent potentials $V(\vec{x})$. Due to the movement of the Earth
our potential is strictly speaking time-dependent but we assume the measurement time to
be small enough to neglect daily and annual modulation effects.

We furthermore assume, for simplicity that the interferometer has a parallelogram shape and the potential
has a constant gradient over the interferometer which is orthogonal to the two sides
$\overline{B_1 C}$ and $\overline{A B_2}$. Under these assumptions
\begin{align}
	\label{eq:Phase_Difference}
	\Delta \Phi = - \frac{1}{\hbar} \oint V(\vec{x}) \diff t = \frac{1}{\hbar} (V_2 - V_1) \, T \;,
\end{align}
with $V_2 = V(\vec{x})$ on the path $\overline{A B_2}$
and $V_1 = V(\vec{x})$ on the path $\overline{B_1 C}$.
This formula is the crucial formula to estimate the size of the effect and 
we discuss the origin of the considered potential in the next section.

We will also consider a different scenario where the two paths can actually be at the
same (or very similar) location but the beam splitter separates the beam into spin-up
and spin-down. Such a setup was discussed, for instance,
in \cite{Parnell:2020wwb} and references therein.
This is also a valid alternative since, as we will see, parts of the potential are spin-dependent
which can lead to a measurable phase difference for spin-up and -down particles.
Even then we can still use the same basic formula from above and, importantly,
this part of the potential is expected to lead to a significantly bigger phase difference.

Here, we discuss two types of matter interferometers relevant for our study. The first one is sensitive
to a spin-independent interaction between the CNB and matter, based on the neutrino-matter potential
important for the MSW effect. Due to the terrestrial CNB
asymmetry on the surface of the Earth, we can place one arm of the interferometer on the Earth's
surface while another arm is positioned underground.
An example for this would be a gradiometer consisting of two atomic
interferometers using, for instance, clouds of  Rubidium atoms. Gradiometers have been employed to
probe the gravitational Aharonov--Bohm
effect~\cite{Overstreet:2021hea} and could be used to probe this terrestrial CNB asymmetry.
In this case, two or more atom interferometers are referenced by a single laser to eliminate some of the laser noise.
Such a setup also allows other
noise elimination such as gravity gradient noise to achieve better sensitivity. Furthermore, the wave
packet separation between the atomic clouds employed in the interferometer can be as large as 25~cm
allowing us to measure the phase difference induced by the CNB between two corresponding interferometer arms.

The second example of a matter interferometer relevant for us is a spin-echo neutron interferometer
which has been employed to probe exotic
spin-dependent interactions between neutron and matter \cite{Parnell:2020wwb}. This kind of instrument utilizes a
beam of neutrons that separates the spin-up and spin-down into two different paths by using magnetic
fields as a beam splitter. The splitting between two neutron beams are controlled by the gradient of the
magnetic field which is directed perpendicularly with respect to the neutron beams. Moreover, another
magnetic field is used to recombine these two beams. Finally, the phase difference experienced by the two
spin states of the neutrons along the two paths is measured. This type of measurement would be
sensitive to a spin-dependent CNB-matter interaction known as the Stodolsky effect. It is possible
that the different spin states of the neutron along the two paths would interact with the CNB resulting
in a non-zero phase shift which could be measured.

While we mentioned here two specific examples for types of matter interferometers, we want to emphasize that our
approach can be easily generalized to other types of matter interferometers as well which might be more
sensitive than the chosen examples. An extensive discussion of different technologies is beyond the scope
of this paper. For a review on atom interferometers,
see, for instance \cite{Cronin:2009zz}.

\section{The effective potentials}
\label{sec:Potentials}

We want to study now the Hamiltonian induced by the CNB on a cloud of atoms in an atomic
interferometer. We focus here solely on fermionic atoms and our calculations also apply to
electron or neutron interferometers.

Since scattering is extremely rare as we will discuss later, we focus mainly on the effective neutrino
potentials induced by the CNB.
This potential is mainly known under two names in the literature, the MSW effect \cite{Wolfenstein:1977ue,Mikheev:1986wj}
and the Stodolsy effect \cite{Stodolsky:1974aq}. As we will see the dominant
effect can be rewritten in terms of a scalar potential, pseudo magnetic fields and a
spin-spin interaction.

Let us begin by looking at the low energy effective Hamiltonian density in the
SM describing neutrino interactions, see, for instance, \cite{Bauer:2022lri},
\begin{align}
	{\cal H}(x) = \frac{G_F}{\sqrt{2}}  \sum_{i,j} \bar{\nu}_i \gamma_\mu (1 - \gamma_5) \nu_j \,  \bar{\psi}_T \gamma^\mu ( V_{ij} - A_{ij} \gamma_5) \psi_T \;,
\end{align}
where we are working in the neutrino mass basis and $\psi_T$ labels here the fermions used in
the considered interferometer.
The matrices $V_{ij}$ and $A_{ij}$ are non-diagonal matrices describing the vector and axial vector coupling
strength of the atom and they also contain flavour effects. The operator here has the structure of a neutral
current interaction. For charged leptons there is also a charged current contribution which can nevertheless
be rewritten into this operator using a Fierz transformation.

We can derive from this Hamiltonian density the potential energy
induced by the CNB background  in the same way as described in \cite{Bauer:2022lri}.
For Dirac neutrinos we find
\begin{align}
	V_\nu^D 
	&= \frac{G_F}{\sqrt{2} } \sum_{k,s_k}  \frac{1}{E_T \, E_k} \Bigl\{  
	(n_k - \bar{n}_k )  V_{kk} (p_k \cdot p_T) 
	- (n_k + \bar{n}_k ) V_{kk} m_k (S_k \cdot p_T) \nonumber\\
	&\phantom{=}  - (n_k - \bar{n}_k )  A_{kk} m_T (p_k \cdot S_T)
	+ (n_k + \bar{n}_k ) A_{kk} m_k \, m_T (S_k \cdot S_T) \Bigr\}
	 \;,
\end{align}
where the sum is over the neutrino mass indices $k$ and neutrino spin-orientations $s_k= \pm 1$.
We label the neutrino number density of the $k$-th mass state with $n_k$ and for anti-neutrinos
with $\bar{n}_k$. The four-vectors
\begin{align}
	\label{eq:Spin_Vector}
	S_i^\mu = s_i \left( \frac{\vec{p}_i \cdot \vec{\mu}_i } {m_i \, |\vec{\mu}_i| } , \frac{\vec{\mu}_i}{|\vec{\mu}_i|} + \frac{ (\vec{p}_i \cdot \vec{\mu}_i) \vec{p}_i  }{ |\vec{\mu}_i| \, m_i (E_i+m_i)} \right) \text{ with } i = k,T \;,
\end{align}
are spin-vectors with $\vec{\mu}_i / |\vec{\mu}_i|$ labelling the direction of the spin in the
rest-frame of the particle and $\vec{\mu}_i$ the magnetic moment of particle $i$. We will use this
expression in our calculations which leads to deviations to the results from, for instance,
\cite{Bauer:2022lri} or \cite{Duda:2001hd} which rewrite the spin-vector as
\begin{align}
	\label{eq:Spin_Helicity}
	S_i^\mu = s_i \left( \frac{|\vec{p}_i|}{m_i} , \frac{E_i}{m_i} \frac{\vec{p}_i}{|\vec{p_i}|} \right) \;, 
\end{align}
which has the advantage that $s_i$ corresponds to helicity and all expressions would just depend
on energy, momenta and masses. But the disadvantage is, that the information about the actual
spin-directions are hidden in the formulas.

For the case of Majorana neutrinos we find for the potential the general expression
\begin{align}
	V_\nu^M
	&= - \frac{G_F}{\sqrt{2} } \sum_{k, s_k}  \frac{2 \, m_k }{E_T \, E_k}  n_k(s_k) \left\{ V_{kk} (S_k \cdot p_T) -  m_T \, A_{kk} (S_k \cdot S_T) \right\} \;.  
\end{align}
Here, we do not introduce $\bar{n}_k$ as there are no anti-neutrinos in the Majorana case. Other
authors use helicity as distinction. We just add an explicit dependence of the neutrino-density
on $s_k$.

Expanding these expressions in terms of energy, momentum, mass and magnetic moment is
lengthy and not very insightful. So let us consider more explicitly the relativistic and the
non-relativistic limit for the neutrinos.

\subsection{Relativistic Limit}

We want to assume now that the fermions in the interferometer all travel in the same direction
with the same velocity which is non-relativistic in the lab-frame. In all the approximations
used the velocities will either be relativistic $|\vec{\beta}| \approx 1$ or they appear to linear
order and after averaging over the velocity distribution
(we label the velocity averages with $\langle \cdot \rangle_v$) we can simply replace the velocities
with their average value, i.e., $\langle \vec{p}_{T} \rangle_v = \vec{p}_{T,0} = m_T \vec{\beta}_{T,0}$.
For the neutrinos we assume in this section that $E_k \gg m_k$ and then on average
$\langle \vec{p}_k \rangle_v \approx E_k \, \vec{e}_{k,0}$.
So the unit vector  $\vec{e}_{k,0}$ labels the direction of the neutrino momentum.

For the relevant averages in the expression for the potentials we then find up to linear
order in the small velocity $\vec{\beta}_{T,0}$
\begin{align}
	 \left\langle \frac{p_k \cdot p_T}{E_T \, E_k}  \right\rangle_v
	 &= 1 - \vec{e}_{k,0} \cdot \vec{\beta}_{T,0}  \;,\\
	m_T \left\langle \frac{p_k \cdot S_T}{E_T \, E_k}  \right\rangle_v
	&\approx 
	s_T \frac{\vec{\mu}_T }{|\vec{\mu}_T|} \cdot (\vec{\beta}_{T,0} - \vec{e}_{k,0} ) \;, \\
	m_k \left\langle \frac{S_k \cdot p_T}{E_T \, E_k}  \right\rangle_v
	&\approx s_k \frac{\vec{e}_{k,0} \cdot \vec{\mu}_k } {|\vec{\mu}_k | }  \, (  1 - 
	 \vec{e}_{k,0} \cdot  \vec{\beta}_{T,0}   ) \;,\\
	m_k \, m_T \, \left\langle \frac{S_k \cdot S_T}{E_T \, E_k}  \right\rangle_v
	  &\approx  s_k \, s_T  \frac{\vec{e}_{k,0} \cdot \vec{\mu}_k } {|\vec{\mu}_k| } 
	  \frac{\vec{\mu}_T \cdot ( \vec{\beta}_{T,0} - \vec{e}_{k,0}  ) } {|\vec{\mu}_T| }
	    \;.
\end{align}
This leads us to the potentials
\begin{align}
	V_{\nu,R}^D &\approx 
		\frac{G_F}{\sqrt{2} } \sum_{k,s_k} 
		 \Biggl\{
		 \left[ (n_k - \bar{n}_k ) 
		- (n_k + \bar{n}_k ) s_k \frac{\vec{e}_{k,0} \cdot \vec{\mu}_k } {|\vec{\mu}_k | } \right] V_{kk} (  1 - \vec{e}_{k,0} \cdot 
	  \vec{\beta}_{T,0}   )  \nonumber\\
		&\phantom{=}  
		- \left[ (n_k - \bar{n}_k )     - 
		 (n_k + \bar{n}_k )  s_k  \frac{\vec{e}_{k,0} \cdot \vec{\mu}_k } {|\vec{\mu}_k| } 
		\right] A_{kk} \, s_T \frac{\vec{\mu}_T }{|\vec{\mu}_T|} \cdot (\vec{\beta}_{T,0} - \vec{e}_{k,0} )
		\Biggr\} \;, \\
	V_{\nu,R}^M &\approx
	 - \sqrt{2} \, G_F \sum_{k, s_k}  n_k(s_k) \, s_k \Biggl\{
	  V_{kk}  \frac{\vec{e}_{k,0} \cdot \vec{\mu}_k } {|\vec{\mu}_k | } (  1 -  \vec{e}_{k,0} \cdot \vec{\beta}_{T,0}  ) \nonumber\\
	  &\quad \quad
	   + A_{kk} s_T  \frac{\vec{e}_{k,0} \cdot \vec{\mu}_k } {|\vec{\mu}_k| } 
	  \frac{\vec{\mu}_T \cdot (\vec{e}_{k,0} - \vec{\beta}_{T,0}) } {|\vec{\mu}_T| } \Biggr\} \;.  
\end{align}
Before we continue we can use another approximation in the relativistic limit, that
is that the magnetic moment of a relativistic particle is approximately aligned
with its momentum. That is we can use $\vec{e}_{k,0} \approx \vec{\mu}_k/|\vec{\mu}_k|$.
In this limit $s_k \, \vec{e}_{k,0} \cdot \vec{\mu}_k / |\vec{\mu}_k |  \to s_k$ and $s_k$ is
identical to helicity. The potentials are then
\begin{align}
	V_{\nu,R}^D &\approx 
		\frac{G_F}{\sqrt{2} } \sum_{k,s_k} 
		 \Biggl\{
		 \left[ (n_k - \bar{n}_k ) 
		- (n_k + \bar{n}_k ) s_k \right] V_{kk} (  1 - \vec{e}_{k,0} \cdot 
	  \vec{\beta}_{T,0}   )  \nonumber\\
		&\phantom{=}  
		- \left[ (n_k - \bar{n}_k )  -  (n_k + \bar{n}_k )  s_k  \right] A_{kk} \, s_T \frac{\vec{\mu}_T }{|\vec{\mu}_T|} \cdot (\vec{\beta}_{T,0} - \vec{e}_{k,0} )
		\Biggr\} \;, \\
	V_{\nu,R}^M &\approx
	 - \sqrt{2} \, G_F \sum_{k, s_k}  n_k(s_k) \, s_k \Biggl\{
	  V_{kk}   ( 1 - 
	  \vec{\beta}_{T,0} \cdot  \vec{e}_{k,0} ) 
	   + A_{kk} s_T 
	  \frac{\vec{\mu}_T \cdot (\vec{e}_{k,0} - \vec{\beta}_{T,0} ) } {|\vec{\mu}_T| } \Biggr\} \;.  
\end{align}

At this point we can identify the matter potential responsible for the MSW effect by averaging over the target magnetic moment
and setting the target velocity to zero. We find
\begin{align}
	V_{\nu,R}^D
	&\to \frac{G_F}{\sqrt{2} } \sum_{k,s_k} 
	(n_k - \bar{n}_k )  V_{kk}  - (n_k + \bar{n}_k ) V_{kk} s_k \approx \sqrt{2} \, G_F \sum_{k} (n_k - \bar{n}_k)  V_{kk} \;, \\
	V_{\nu,R}^M
	&\to - \sqrt{2} \, G_F \sum_{k,s_k} n_k(s_k)  s_k V_{kk} \;.  
\end{align}
For the Dirac scenario we have used that $n_k(s_k = +1) \approx \bar{n}_k(s_k = -1)$.
We will comment on this in more detail in Sec.~\ref{sec:Results}.
These are indeed the standard expressions for the MSW potential but now for matter
in a neutrino background. There is also no difference between the Dirac and Majorana case
if we identify $n_k = n_k(-1)$ and $\bar{n}_k=n_k(+1)$ as expected from the
practical Majorana-Dirac confusion theorem~\cite{Kayser:1982br}.

We can also write the potentials in a suggestive form
\begin{align}
	\label{eq:VnuR_phi_B}
	V_{\nu,R} = \phi_{\nu,R}(\vec{r}) - \vec{\mu}_T \cdot \vec{B}_{\nu,R} (\vec{r}) \;,
\end{align}
with the scalar potentials 
\begin{align}
	\phi_{\nu,R}^D(\vec{r}) &= 
	\frac{G_F}{\sqrt{2} } \sum_{k,s_k} 
		 \left[ (n_k - \bar{n}_k )  - (n_k + \bar{n}_k )  s_k \right] V_{kk}  (1 - \vec{e}_{k,0} \cdot \vec{\beta}_{T,0})  \;, \\
	\phi_{\nu,R}^M(\vec{r}) &= - \sqrt{2} \, G_F \sum_{k,s_k} n_k(s_k) s_k V_{kk}   ( 1 -  \vec{e}_{k,0} \cdot \vec{\beta}_{T,0}  ) \;,
\end{align}
and the pseudo magnetic fields coupling to a magnetic moment
\begin{align}
	\vec{B}_{\nu,R}^D  &=  \frac{G_F}{\sqrt{2} } \frac{s_T }{|\vec{\mu}_T|}  \sum_{k,s_k}  [  
	 (n_k - \bar{n}_k )   - (n_k + \bar{n}_k )  s_k  ] A_{kk} ( 
	    \vec{\beta}_{T,0}  -   \vec{e}_{k,0} ) 
	 \;, \\
	\vec{B}_{\nu,R}^M  &=  - \sqrt{2} \, G_F  \frac{s_T}{|\vec{\mu}_T|}  \sum_{k,s_k} n_k(s_k) s_k A_{kk}  ( 
	    \vec{\beta}_{T,0} -  \vec{e}_{k,0} ) \;,
\end{align}
for the Dirac and Majorana case, respectively. The position dependence of the scalar
potential and the pseudo magnetic field originate from the position dependence of the
neutrino number densities. 

The Stodolsky effect is the part of the potential which depends on the spin of the target
and which we have rewritten here in terms of a pseudo $\vec{B}$-field. At this point though
we want to stress more explicitly that these are not real $\vec{B}$-fields related to electromagnetism.
They completely originate from weak interactions. This notation though makes
a comparison to the interferometer literature or other detection schemes far more convenient.

One way to tell that the $\vec{B}$-fields are not magnetic in nature is also apparent from the
fact that they do not only depend on properties of the neutrino background (their source) but also on
$|\vec{\mu}_T|$ and $A_{kk}$ which contain properties of the probe. So the same background would
lead to different effective $\vec{B}$-fields depending on what is used in the interferometer.

\subsection{Non-relativistic limit}

We now move on to the non-relativistic limit which is well motivated since at least part of the CNB today
should be non-relativistic. The results in this section can also be applied straight-forwardly to fermionic
DM coupling via vectors or axial-vectors to ordinary matter. For other operators and an extensive
discussion of the Dark Stodolsky effect, see~\cite{Rostagni:2023eic}.

Compared to the previous case we now assume for the neutrinos that they have an average
momentum $\vec{p}_{k,0} = m_k \, \vec{\beta}_{k,0}$ with $|\vec{p}_{k,0}| \ll E_k$.
Then we find for the relevant velocity averages up to linear order in small velocities
\begin{align}
	 \left\langle \frac{p_k \cdot p_T}{E_T \, E_k}  \right\rangle_v
 	&\approx 1 
	\;,\\
	m_T \left\langle \frac{p_k \cdot S_T}{E_T \, E_k}  \right\rangle_v
	&\approx s_T \frac{\vec{\mu}_T }{|\vec{\mu}_T|} \cdot (\vec{\beta}_{T,0} - \vec{\beta}_{k,0} ) \;, \\
	m_k \left\langle \frac{S_k \cdot p_T}{E_T \, E_k}  \right\rangle
	&\approx s_k  \frac{\vec{\mu}_k }{|\vec{\mu}_k|} \cdot  (\vec{\beta}_{k,0} - \vec{\beta}_{T,0} )  \;,\\
	m_k \, m_T \, \left\langle \frac{S_k \cdot S_T}{E_T \, E_k}  \right\rangle_v
	&\approx
	- s_k s_T  \frac{\vec{\mu}_T }{|\vec{\mu}_T|}  \cdot  
	\frac{\vec{\mu}_k}{|\vec{\mu}_k|}   \;.
\end{align}
Then we find for the potentials up to the same order 
\begin{align}
	\label{eq:Vnu_NR_D}
	V_{\nu,NR}^D  
	&\approx \frac{G_F}{\sqrt{2} } \sum_{k,s_k} \Biggl\{  
	(n_k - \bar{n}_k )  V_{kk} 
	- s_k (n_k + \bar{n}_k ) V_{kk}  \frac{\vec{\mu}_k }{|\vec{\mu}_k|} \cdot  (\vec{\beta}_{k,0} - \vec{\beta}_{T,0} )   \nonumber\\
	&\phantom{=}  - s_T (n_k - \bar{n}_k )  A_{kk}  \frac{\vec{\mu}_T }{|\vec{\mu}_T|} \cdot (\vec{\beta}_{T,0} - \vec{\beta}_{k,0} )
	- s_k \, s_T (n_k + \bar{n}_k ) A_{kk}  \frac{\vec{\mu}_T }{|\vec{\mu}_T|} \cdot \frac{\vec{\mu}_k}{|\vec{\mu}_k|}  \Biggr\} \;, \\
	\label{eq:Vnu_NR_M}
	V_{\nu,NR}^M
	&\approx  \sqrt{2} \, G_F \sum_{k,s_k} n_k s_k \Biggl\{  V_{kk} \frac{\vec{\mu}_k }{|\vec{\mu}_k|} \cdot  (\vec{\beta}_{k,0} - \vec{\beta}_{T,0} )  -  s_T \, A_{kk} \frac{\vec{\mu}_T }{|\vec{\mu}_T|} \cdot \frac{\vec{\mu}_k}{|\vec{\mu}_k|}  \Biggr\} \;.
\end{align}
Latest here for the Majorana case it becomes apparent how our expressions differ from the result from \cite{Bauer:2022lri} 
or \cite{Duda:2001hd} (apart from a different approach to take the velocity averages). In our expression it is apparent that the
axial vector coupling for the Majorana case leads to a pure spin-spin interaction while in \cite{Bauer:2022lri,Duda:2001hd} there
is no neutrino spin vector in their final expressions. For the relativistic case there is also a contribution from the spin-spin
interaction which is nevertheless hidden even in our approach since in that limit $\vec{\mu}_k / |\vec{\mu}_k| \approx \vec{e}_{k,0}$.

Here we want to again make an analogy to scalar potentials and $\vec{B}$-fields but there is also additionally
the aforementioned spin-spin interaction term
\begin{align}
	\label{eq:VnuNR_phi_B}
	V_{\nu,NR} = \phi_{\nu,NR}(\vec{r}) - \vec{\mu}_T \cdot \vec{B}_{\nu,NR} (\vec{r}) - \sum_{k,s_k} \vec{\mu}_k \cdot \vec{B}_{T,NR} (\vec{r}) - \sum_{k,s_k} I_{k,NR}(\vec{r}) \, \vec{\mu}_k \cdot \vec{\mu}_T  \;.
\end{align}
Since the magnetic moment of the neutrinos do not have to align with the momentum anymore we
can now also define a $\vec{B}$-field induced by the target atoms on the neutrinos and a coupling
of the magnetic moments on each other.

The relevant quantities are then for the Dirac case
\begin{align}
	\phi_{\nu,NR}^D(\vec{r}) &=   \frac{G_F}{\sqrt{2}} \sum_{k, s_k}
	(n_k - \bar{n}_k )  V_{kk}  
	\;, \\
	\vec{B}_{\nu,NR}^D (\vec{r}) &=  \frac{G_F}{\sqrt{2} } \sum_{k,s_k}  (n_k - \bar{n}_k)  \frac{ s_T \, A_{kk} }{|\vec{\mu}_T|}
	 (\vec{\beta}_{T,0}  -   \vec{\beta}_{k,0}  ) 
	 \;, \\
	\vec{B}_{T,NR}^D (\vec{r}) &=  \frac{G_F}{\sqrt{2} } \sum_{k,s_k}
	 (n_k + \bar{n}_k ) \frac{s_k \, V_{kk} }{|\vec{\mu}_k|} (\vec{\beta}_{k,0} - \vec{\beta}_{T,0} ) \;, \\
	I_{k,NR}^D(\vec{r}) &= \frac{G_F}{\sqrt{2}} \sum_{k,s_k}  (n_k + \bar{n}_k) \frac{s_k \, s_T \, A_{kk}}{ |\vec{\mu}_k|  \, |\vec{\mu}_T|} \;,
\end{align}	
and for the Majorana case
\begin{align}
	\phi_{\nu,NR}^M(\vec{r}) &= 0 \;, \\
	\vec{B}_{\nu,NR}^M (\vec{r}) &=  0 \;, \\
	\vec{B}_{T,NR}^M (\vec{r}) &=  \sqrt{2} \, G_F \sum_{k,s_k} n_k(s_k)  \frac{ s_k \, V_{kk} }{|\vec{\mu}_k|}    (\vec{\beta}_{T,0} - \vec{\beta}_{k,0} )  \;, \\
	I_{k,NR}^M(\vec{r}) &= \sqrt{2} \, G_F \sum_{k,s_k}  n_k(s_k)  \frac{s_k \,  s_T \, A_{kk}}{ |\vec{\mu}_k|  \, |\vec{\mu}_T|} \;.
\end{align}
We will discuss some estimates for the size of these quantities later. 

\section{Results}
\label{sec:Results}

In the following we want to present some numerical results for the generic formulas in the previous
sections and compare them to current and future experimental sensitivities. We can split the results into two main categories,
the neutrino matter potential (from the vector interaction) and the Stodolsky potential (from the axial-vector interaction). We will also provide some estimates for the scalar potentials,
pseudo $B$-fields, the spin-spin interaction, DM and the case of scattering.
But first, we comment on some simplifying assumptions which we will use throughout this section.

In this work we present results for an electron
interferometer, a neutron interferometer and atom interferometers using $^{87}$Sr  as being considered,
for instance, in AION~\cite{Badurina:2019hst}. There are also other proposals using $^{87}$Rb like 
BECCAL~\cite{Aveline:2020kla,Frye:2021jgc,Elliott:2018cal} but an $^{87}$Rb atom 
is a boson so our formulas do not apply to this 
case.
In that case the standard Stodolsky effect coupling to the magnetic moment would be
absent but the neutrino matter potential would still be present. To compare the size of the effects, we focus on interferometers using fermions. 
We will see that the Stodolsky effect is expected to be more sizeable.
For simplicity, we also do not discuss interferometers using molecules.

In our conventions we have for the electron, cf.~\cite{Bauer:2022lri},
\begin{align}
	V^e_{kk} = -1/2 + 2 \, \sin^2 \theta_W + |U_{ek}|^2 \text{ and } A^e_{kk} = -1/2 + |U_{ek}|^2 \;.
\end{align}
For neutrons and protons we have instead
\begin{align}
	V^n_{kk} = -1/2  &\text{ and } A^n_{kk} = -1/2 \;, \\
	V^p_{kk} = +1/2 - 2 \, \sin^2 \theta_W  &\text{ and } A^p_{kk} = +1/2 \;.
\end{align}
Hence, for an arbitrary atom with $Z$ protons, $N$ neutrons and $E$ electrons
we then have
\begin{align}
	V^{\text{atom}}_{kk} &= \frac{1}{2} (Z-N-E) + 2 \, \sin^2 \theta_W (E-Z) + E \, |U_{ek}|^2 \;, \\
	 A^{\text{atom}}_{kk} &= \frac{1}{2} (Z - N - E) + E \, |U_{ek}|^2 \;.
\end{align}
These formulas would also apply to a single neutron or electron.
For the numerical results we use $|U_{e1}|^2 \approx 0.823$~\cite[IC24, Normal Ordering, best-fit value]{Esteban:2024eli}
which appears in $V_{11}$ and $A_{11}$.
For the other two relevant neutrino mixing matrix elements which will appear together we can use the
 unitarity of the mixing matrix,
i.e., $|U_{e2}|^2 + |U_{e3}|^2 = 1 - |U_{e1}|^2$.

So for neutral $^{87}$Sr with $Z=38$, $N = 49$ and $E=38$
\begin{align}
	V^{\text{87Sr}}_{kk} &= - \frac{49}{2}  + 38 \, |U_{ek}|^2 \;, \\
	 A^{\text{87Sr}}_{kk} &= - \frac{49}{2} + 38 \, |U_{ek}|^2 \;.
\end{align}
At this point, we need to discuss coherence factors in a bit more detail. Low-energy neutrinos
have rather long wave-lengths, for the CNB the typical de-Broglie wavelength is $\mathcal{O}(0.1\text{ cm})$,
see, for instance, the discussion in \cite{Domcke:2017aqj} and references therein. That justifies why we included 
here all nucleons
and the electrons in the shell of the atoms which applies to vector and axial-vector currents,
see~\cite{Freedman:1973yd}. So we treat the atoms as a single unit with a single wave function.

One might wonder at this point, if the coherence should be applied to multiple atoms since the neutrino 
wavefunction can be bigger than the inter-atom spacing.
Our current understanding is, that this is not quite correct since the interferometer reads out the internal
states of the atoms in the interferometer. So if one would want to boost the effect using coherence
one would have to consider ``bigger'' states like molecules, but we will not consider explicit examples here
or include a coherence factor over multiple atoms.

The next thing which we need to discuss are the values for the neutrino number densities.
In \cite{Long:2014zva}, for instance, they argue that for Dirac neutrinos (translated into our notation)
\begin{align}
	n_k(s_k = -1) &= n_0  = \bar{n}_k(s_k = +1) \;, \\
	n_k(s_k = +1) &\approx 0 \approx \bar{n}_k(s_k = -1)\;,
\end{align}
while for Majorana neutrinos they find
\begin{align}
	n_k(s_k = -1) &= n_0  = n_k(s_k = +1) \;,
\end{align}
where $n_0 \approx 56 \text{ cm}^{-3}$ as already mentioned and they neglect the possibility of having a
lepton asymmetry which is crucial for us here. So we will assume here instead that some mechanism created a
sizeable lepton asymmetry in the early universe such that
for Dirac neutrinos 
\begin{align}
	n_k(s_k = -1) - \bar{n}_k(s_k = +1) &= \frac{\Delta_\nu}{\text{cm}^3} \;, \\
	n_k(s_k = +1) \approx 0 \approx \bar{n}_k(s_k = &-1)\;,
\end{align}
and for Majorana neutrinos
\begin{align}
	n_k(s_k = -1) - n_k(s_k = +1) &= \frac{\Delta_\nu}{\text{cm}^3} \;.
\end{align}
While in the SM we would expect $|\Delta_\nu| \ll 1$, experimentally very large values
$|\Delta_\nu| \lesssim 35$ are allowed \cite{Domcke:2025jiy}. Note that for the sake of simplicity
we treat the asymmetry as flavor universal.

In our analysis, we will also neglect the dependence of the potentials on the target velocity $\vec{\beta}_{T,0}$
as it is usually very small, much smaller than the velocity of the neutrinos with $|\vec{\beta}_{k,0}| \gtrsim 10^{-4}$.
Actually, for the non-relativistic neutrinos, we will consider a universal value of
$|\vec{\beta}_{k,0}| \equiv |\vec{\beta}_{0}| \approx 7.7 \times 10^{-4}$ 
for our numerical results which is approximately the virial velocity of our galaxy at
the Sun's location. For relativistic neutrinos we have already used that
$|\vec{\beta}_{k,0}| \approx 1$.

The relativistic species is also expected to have a different velocity direction compared to the
non-relativistic species. For the non-relativistic species, we set their velocity direction equal to the virial velocity direction
which is roughly in the direction of the Cygnus constellation which is in galactic coordinates
$(l,b) \approx (270^\circ,0^\circ)$. For a relativistic relic neutrino species we assume the direction is aligned with the CMB dipole
moment which is directed towards galactic coordinates $(l,b) \approx (264^{\circ},48^{\circ})$. So their
relative angle
\begin{align}
	\label{eq:Relative_Velocity_Angle}
	\cos \theta_v \approx \cos 48^\circ \approx 0.67 \;.
\end{align}

Collecting all these assumptions, we find that for the CNB induced potential for one relativistic
species of neutrinos labeled $k$ is
\begin{align}
	\label{eq:VnuRD}
	V_{k,R}^D &\approx \sqrt{2}
		 \, G_F \frac{\Delta_\nu}{\text{cm}^3}  
		 \Biggl\{
		 V_{kk} + s_T \, A_{kk}  \frac{\vec{\mu}_T }{|\vec{\mu}_T|} \cdot \vec{e}_{k,0} 
		  \Biggr\} \equiv V_{k,R}
	\;, \\
	\label{eq:VnuRM}
	V_{k,R}^M &\approx
	 \sqrt{2} \, G_F \frac{\Delta_\nu}{\text{cm}^3} \Biggl\{
	  V_{kk} 
	   + s_T \, A_{kk} 
	  \frac{\vec{\mu}_T} {|\vec{\mu}_T| } \cdot \vec{e}_{k,0} \Biggr\} \equiv V_{k,R} \;,  
\end{align}
where we can see the practical Majorana-Dirac confusion theorem~\cite{Kayser:1982br} in action.

For the non-relativistic case we still need to discuss what we assume for $\vec{\mu}_k/ |\vec{\mu}_k|$,
that is the direction of the neutrino magnetic moment.
In principle, due to gravitational and electromagnetic interactions the neutrino magnetic moment might be misaligned
with the momentum. 
But such distortions are likely stochastic on the way from the early universe until today and we assume
$\vec{\mu}_k/ |\vec{\mu}_k| \approx \vec{\beta}_{k,0}/ |\vec{\beta}_{k,0}|$ which is consistent with using
the helicity spin-vector, eq.~\eqref{eq:Spin_Helicity}. 
Then we find for a non-relativistic neutrino 
\begin{align}
	\label{eq:VnuNRD}
	V_{k,NR}^D  
	&\approx \frac{G_F}{\sqrt{2}} \frac{\Delta_\nu}{\text{cm}^3} \Biggl\{  
	V_{kk}   +  s_T \, A_{kk}  \frac{\vec{\mu}_T }{|\vec{\mu}_T|} \cdot  \frac{\vec{\beta}_{k,0}}{ |\vec{\beta}_{k,0}| } 
	\Biggr\}  ( 1 +  |\vec{\beta}_{k,0}|  ) \;, \\
	\label{eq:VnuNRM}
	V_{k,NR}^M
	&\approx  \sqrt{2} \, G_F \frac{\Delta_\nu}{\text{cm}^3} \Biggl\{   V_{kk} |\vec{\beta}_{k,0}|  +  s_T \, A_{kk} \frac{\vec{\mu}_T }{|\vec{\mu}_T|} \cdot \frac{\vec{\beta}_{k,0}}{ |\vec{\beta}_{k,0}| }  \Biggr\}  \;,
\end{align}
where there is clearly a difference for Dirac and Majorana neutrinos.

There is one thing we did not discuss yet in detail and that is the question which neutrinos are relativistic or non-relativistic.
Since we do not know the neutrino mass scale and ordering (yet) we have to make an assumption here as well. To get some idea
about the size of the effects we will consider two scenarios.

The first scenario is where we assume normal ordering of neutrinos and the lightest neutrino to be
approximately massless and relativistic
while the other two neutrinos are non-relativistic today.
So for our numerical
results we will consider for the relativistic case the potential energies
\begin{align}
	E_R^D &= \sqrt{2} \, G_F \frac{\Delta_\nu}{\text{cm}^3}  
		 \Biggl\{
		 V_{11} + s_T \, A_{11}  \frac{\vec{\mu}_T }{|\vec{\mu}_T|} \cdot \vec{e}_{1,0}  
		  \Biggr\}  \nonumber\\
		  &\phantom{=} + \frac{G_F}{\sqrt{2}} \frac{\Delta_\nu}{\text{cm}^3} \sum_{k=2}^3 \Biggl\{  
	 V_{kk}    +  s_T \, A_{kk}  \frac{\vec{\mu}_T }{|\vec{\mu}_T|} \cdot \frac{\vec{\beta}_{k,0}}{ |\vec{\beta}_{k,0}| }
	  \Biggr\} ( 1 +  |\vec{\beta}_{k,0}|  ) \;,\\
	E_R^M &= \sqrt{2} \, G_F \frac{\Delta_\nu}{\text{cm}^3}  
		 \Biggl\{
		 V_{11} + s_T \, A_{11}  \frac{\vec{\mu}_T }{|\vec{\mu}_T|} \cdot  \vec{e}_{1,0}  
		  \Biggr\}  \nonumber\\
		  &\phantom{=} + \sqrt{2} \, G_F \frac{\Delta_\nu}{\text{cm}^3} \sum_{k=2}^3 \Biggl\{   V_{kk} |\vec{\beta}_{k,0}|  +  s_T \, A_{kk} \frac{\vec{\mu}_T }{|\vec{\mu}_T|} \cdot \frac{\vec{\beta}_{k,0}}{ |\vec{\beta}_{k,0}| }  \Biggr\}   \;.
\end{align}

We also consider a scenario where all neutrinos are non-relativistic today. In this case the expressions for the potential
energies are just
the results in eqs.~\eqref{eq:VnuNRD} and \eqref{eq:VnuNRM} where we sum $k$ from $1$ to $3$,
\begin{align}
	E_{NR}^D = \sum_{k=1}^3 V_{k,NR}^D \;,\\
	E_{NR}^M = \sum_{k=1}^3 V_{k,NR}^M \;.
\end{align}

\subsection{Neutrino-Matter Potential}

For the neutrino-matter potential related to the MSW effect, we will consider an unpolarized target. In other words, when averaging over the matter ensemble in the interferometers
the parts proportional to $s_T \, \vec{\mu}_T  / |\vec{\mu}_T|$ average to zero. So the question is how one could actually measure any effect then,
since the potentials seem to be constant. For this question we refer to a recent series of papers  
\cite{Arvanitaki:2022oby,Huang:2024tog,Gruzinov:2024ciz,Kalia:2024xeq} which argue that due to diffraction effects there will be an additional
neutrino asymmetry, $r_\nu$, generated near the surface of the Earth. Although the numbers slightly differ depending on the assumptions,
they find an asymmetry of order of $\mathcal{O}(10^{-8})$ which is also in general flavor dependent.
Again, to simplify the discussion
we just introduce here one flavor universal parameter.

\begin{figure}
	\centering
	\includegraphics[width=\textwidth]{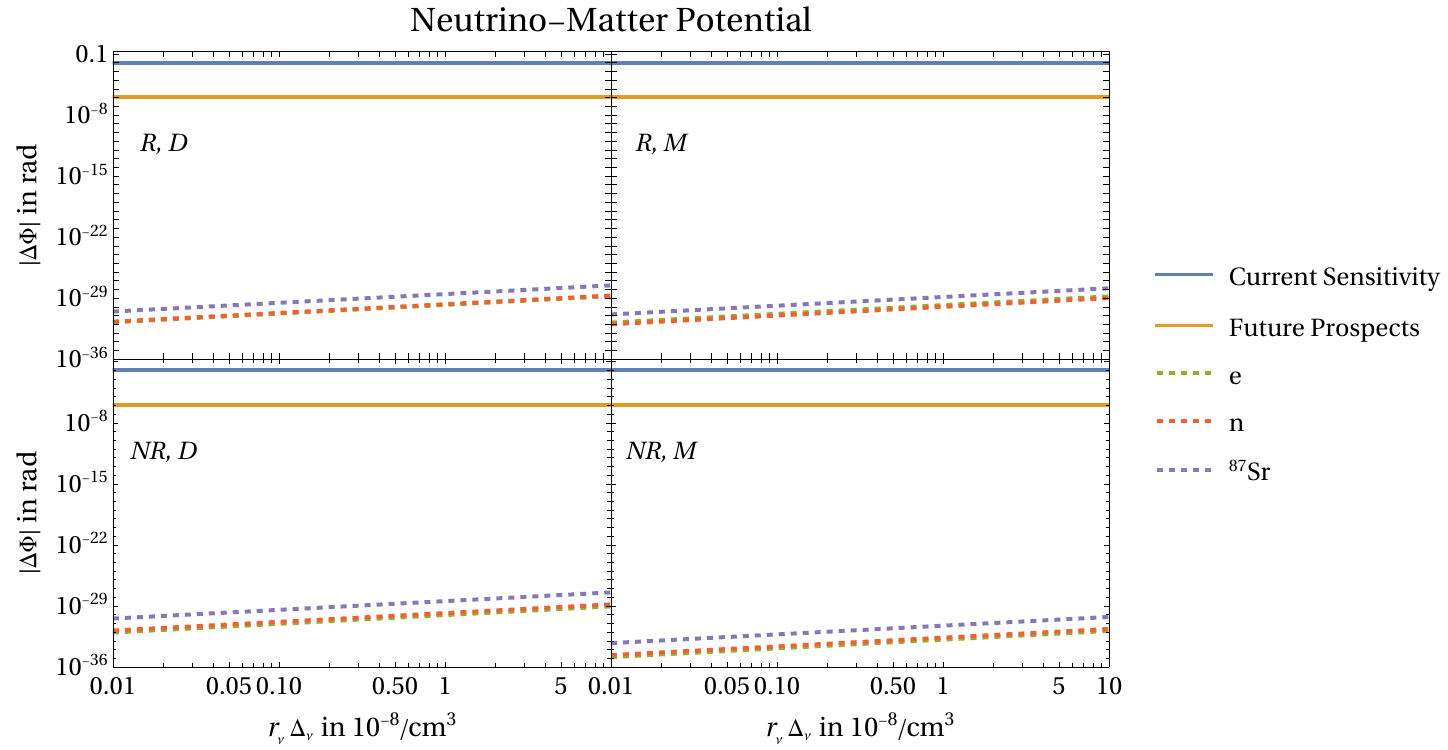}
	\caption{\label{fig:MSW}
		Dependence of $|\Delta \Phi|$ in an electron, neutron and $^{87}$Sr 
		interferometer with $T = 1$~s for the four cases we discuss. Here we plot the effect of the neutrino-matter potential depending on
		the combined parameter $r_\nu \, \Delta_\nu$ in $10^{-8}/$cm$^3$.
		The electron and neutron lines almost overlap.
		The current sensitivity taken here is 0.01~rad inspired from results on measuring the
		gravitational Aharonov--Bohm effect~\cite{Overstreet:2021hea}. In the future atom
		interferometers might be able to resolve phase shifts down to about 
		$10^{-6}$~rad~\cite{Aveline:2020kla,Frye:2021jgc,Elliott:2018cal,Du:2022ceh}
		(which we show as future prospects).
		For more details, see main text.
	}
\end{figure}

We then assume that one of the interferometer arms is in a region where the asymmetry is enhanced by diffraction effects and in the other arm
there is no enhancement or it has the opposite sign.
We then arrive at rather simple approximate formulas for the phase difference from eq.~\eqref{eq:Phase_Difference} which is
for the case with one relativistic neutrino
\begin{align}
	\Delta \Phi^{R,D}_{\text{MSW}} &\approx \frac{T}{\hbar} \frac{G_F}{\sqrt{2}} \frac{r_\nu \, \Delta_\nu}{\text{cm}^3}  
		 \Bigl\{
		 2 \, V_{11} + (V_{22} + V_{33}) ( 1 +  |\vec{\beta}_{0}|  )  
		  \Bigr\} \nonumber\\
		  &\approx 1.61 \times 10^{-30} \frac{T}{1\text{ s}} \frac{r_\nu}{10^{-8}} \frac{\Delta_\nu}{\text{cm}^3}    (E - 1.20 \, N + 0.09 \, Z)  \;, \\
	\Delta \Phi^{R,M}_{\text{MSW}} &\approx \frac{T}{\hbar} \sqrt{2 } \, G_F  \frac{r_\nu \, \Delta_\nu}{\text{cm}^3}  
		 \Bigl\{
		 V_{11} + (V_{22} + V_{33})  |\vec{\beta}_{0}|  
		  \Bigr\} \nonumber\\
		  &\approx 1.51 \times 10^{-30} \frac{T}{1\text{ s}}  \frac{r_\nu}{10^{-8}} \frac{\Delta_\nu}{\text{cm}^3}  (E - 0.64 \, N + 0.05 \, Z) \;,
\end{align}
and for the case with only non-relativistic neutrinos
\begin{align}
	\Delta \Phi^{NR,D}_{\text{MSW}} &\approx \frac{T}{\hbar} \frac{G_F}{\sqrt{2}} \frac{r_\nu \, \Delta_\nu}{\text{cm}^3} \sum_{k=1}^3 
	 V_{kk}  ( 1 +  |\vec{\beta}_{0}|  )  \nonumber\\
	 &\approx 8.56 \times 10^{-31} \frac{T}{1\text{ s}} \frac{r_\nu}{10^{-8}} \frac{\Delta_\nu}{\text{cm}^3}   (E - 1.69 \, N + 0.13 \, Z) 
	   \;, \\
	\Delta \Phi^{NR,M}_{\text{MSW}} &\approx  \frac{T}{\hbar} \sqrt{2} \, G_F \frac{r_\nu \, \Delta_\nu}{\text{cm}^3} \sum_{k=1}^3 
	 V_{kk}    |\vec{\beta}_{0}|   \nonumber\\
	 &\approx 1.31 \times 10^{-33} \frac{T}{1\text{ s}}  \frac{r_\nu}{10^{-8}} \frac{\Delta_\nu}{\text{cm}^3}   (E - 1.69 \, N + 0.13 \, Z) 
	   \;.
\end{align}

We have plotted the above numerical approximations in Fig.~\ref{fig:MSW} as a function
of $r_\nu \, \Delta_\nu$ for the different cases considered and compared them to current and future
experimental sensitivities. As current sensitivity we consider a benchmark value of
0.01~rad inspired from results on measuring the gravitational Aharonov--Bohm 
effect~\cite{Overstreet:2021hea}. In the future, atom
interferometers might be able to resolve phase shifts down to about 
$10^{-6}$~rad~\cite{Aveline:2020kla,Frye:2021jgc,Elliott:2018cal,Du:2022ceh}
(which we show as future prospects).

Unfortunately, comparing these numbers to our numerical formulas from above or as
displayed in Fig.~\ref{fig:MSW} even the most ambitious proposals will probably not be
sensitive enough to measure the CNB directly. Our benchmark values here fall short
by at least about ten orders of magnitude. But we want to emphasize that we have made somewhat
conservative assumptions here. New physics interactions between neutrinos and matter can
potentially increase the low energy $G_F$ and the appropriate value for $r_\nu$.
There could
also be progress in matter interferometers which allow to measure molecules instead of atoms
which can lead to some more enhancements.

One simple estimate for the sensitivity of a matter interferometer is given by
\begin{align}
	\label{eq:delta_Delta_Phi}
	\delta (\Delta \Phi) = \frac{1}{\sqrt{N_c}} \;,
\end{align}
with $N_c$ the number of atoms in the cloud in the interferometer. In the most optimistic scenario
one would have to reach here $\delta (\Delta \Phi) \sim 10^{-27}$ and one would need $10^{55}$ Strontium
atoms which is huge.

It will also be important to understand the neutrino mass scale and the nature of neutrinos (Dirac vs.\ Majorana).
In our simple estimates this can make a difference in terms of expectations of three orders of
magnitude.

\subsection{Stodolsky Potential}

The second effect we consider is using interferometers where the two arms are not necessarily separated
spatially, but instead we assume 
that the beam splitters separate/merge the spin of the matter in the interferometer as it was discussed
for neutrons, for instance, in~\cite{Parnell:2020wwb} where they reached a sensitivity of the order of
a few $10^{-3}$ rad. We also found another example \cite{Cimmino:1989zz} where they measured the
Aharonov--Casher effect
which depends on the magnetic moment of the neutron. They utilized about $2.5\times 10^{7}$ neutrons
and reached also a sensitivity at the order of $10^{-3}$ rad but in their setup they did not separate
spin-up from spin-down neutrons so this just serves as another example how in interferometers
one can measure spin-dependent effects.

Here in this section we will neglect the MSW effect discussed previously, i.e., we assume the arms are not separated enough to
notice any diffraction effects. As a consequence only the parts with the magnetic moment of the targets survive.
To get some estimate
of the maximal effect size we also want to assume that the the magnetic moments of the targets 
are aligned with the direction of the non-relativistic neutrino species. 
As we mentioned we expect there to be an angle between the velocities of the relativistic and non-relativistic
neutrino species, see, eq.~\eqref{eq:Relative_Velocity_Angle}. Furthermore, we assume that
the two interferometer arms are completely polarized in opposite directions.

We can then again find some rather simple expressions for the phase
differences. First for the case with one relativistic neutrino
\begin{align}
	\Delta \Phi^{R,D}_{\text{ST}} &\approx \frac{T}{\hbar}  \sqrt{2} \, G_F \frac{\Delta_\nu}{\text{cm}^3}  
		 \Biggl\{
		  2 \, A_{11} \, \cos \theta_v  + (A_{22} + A_{33}) ( 1 +  |\vec{\beta}_{0}|  ) 
		  \Biggr\}
		    \nonumber\\
		  &\approx - 7.60  \times 10^{-23} \frac{T}{1\text{ s}}   \frac{\Delta_\nu}{\text{cm}^3}  (E + 4.23 \, N - 4.23 \, Z)   \;,\\
	\Delta \Phi^{R,M}_{\text{ST}} &\approx \frac{T}{\hbar} \sqrt{2} \, G_F \frac{\Delta_\nu}{\text{cm}^3}  
		 \Biggl\{
		   2 \, A_{11} \,  \cos \theta_v + 2 \, (A_{22} + A_{33})
		   \Biggr\}   
		   \nonumber\\
		  &\approx -  2.35  \times 10^{-22} \frac{T}{1\text{ s}}   \frac{\Delta_\nu}{\text{cm}^3}   (E + 2.19 \, N - 2.19 \, Z) \;,
\end{align}
and for the case with only non-relativistic neutrinos
\begin{align}
	\Delta \Phi^{NR,D}_{\text{ST}} &\approx \frac{T}{\hbar}  \sqrt{2} \, G_F \frac{\Delta_\nu}{\text{cm}^3} \sum_{k} 
	   A_{kk}
	   ( 1 +  |\vec{\beta}_{0}|  ) 
	\nonumber\\
		&\approx - 9.66  \times 10^{-23} \frac{T}{1\text{ s}}   \frac{\Delta_\nu}{\text{cm}^3}   (E + 3 \, N - 3 \, Z) \;,\\
	\Delta \Phi^{NR,M}_{\text{ST}} &\approx \frac{T}{\hbar} \sqrt{2} \, G_F \frac{\Delta_\nu}{\text{cm}^3} \sum_{k}  2 \, A_{kk}   \nonumber\\
		&\approx - 1.93  \times 10^{-22} \frac{T}{1\text{ s}}   \frac{\Delta_\nu}{\text{cm}^3}  (E + 3 \, N - 3 \, Z) \;.
\end{align}

\begin{figure}
	\centering
	\includegraphics[width=\textwidth]{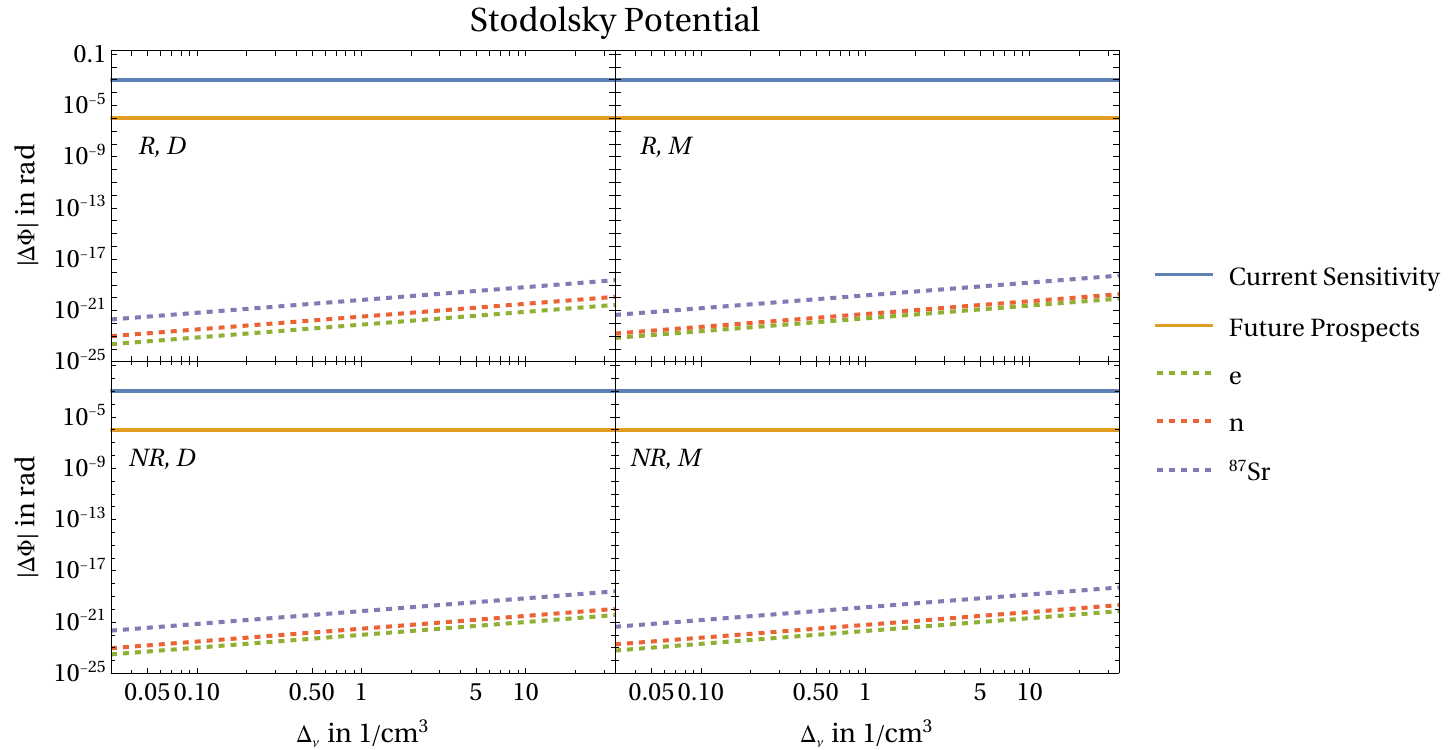} 
	\caption{\label{fig:ST}
		Dependence of $|\Delta \Phi|$ in an electron, neutron and $^{87}$Sr 
		interferometer with $T = 1$~s for the four cases we discuss. Here we plot the effect of the Stodolsky potential depending on
		the parameter $\Delta_\nu$ in $1/$cm$^3$. 
		The current sensitivity taken here is $10^{-3}$~rad inspired from current results from the spin-echo
		neutron interferometry result \cite{Parnell:2020wwb}. For the future prospects we take the
		same number as in Fig.~\ref{fig:MSW}. 
		For more details, see main text.
	}
\end{figure}

We have plotted the above numerical approximations in Fig.~\ref{fig:ST} as a function
of $\Delta_\nu$ for the different cases considered. The immediate thing to be noticed is that the
expected phase differences here are significantly larger than for the neutrino-matter potential.
This is not surprising since for the neutrino-matter potential the effect is suppressed by the small
gradient induced by diffraction. Here, the phase difference is driven by the energy splitting
between the two spin-orientations of the matter in the interferometer.

As a consequence this scenario is much more promising and the expectations are only about eleven orders
of magnitude below the most optimistic future prospects. But as a caveat, these prospects are taken from
proposals which are not discussing beam splitters which separate spin polarizations for which we could
not find prospects for the future.

Again, the same caveats as before apply and new physics effects as well as interferometers with a
significantly larger $N_c$ might boost the expected signal into an observable region. Using eq.~\eqref{eq:delta_Delta_Phi}
we would need here ``only'' about $10^{36}$ atoms which is much more feasible than the previous case but still
very far fetched.
A better
understanding of the properties of neutrinos would again allow us to refine the expectations for future
experiments.

\subsection[Scalar Potentials, Pseudo \texorpdfstring{$B$}{B}-fields and Spin-Spin Interactions]{Scalar Potentials, Pseudo \texorpdfstring{$\boldsymbol{B}$}{B}-fields and Spin-Spin Interactions}

In the previous section we have quantified the phase difference induced by the effective potential.
And as we have mentioned before one can split this effective potential into different parts which have
their own physical interpretations. To be precise, we have identified a scalar potential, potentials
describing a pseudo magnetic field coupling to the matter and neutrino spin respectively, and a spin-spin
interaction potential. While the phase difference is the relevant observable for interferometers we will
discuss here briefly some numerical values for the different parts of the effective potential which might
be interesting for other experimental setups and might be easier to compare and use in estimates.

\subsubsection{Relativistic Case}

Let us begin with the simpler case for a relativistic neutrino species labeled $k$ and
as target we will always assume here a single neutron for better comparison.
As displayed in eq.~\eqref{eq:VnuR_phi_B} we can reinterpret
the potential in terms of a scalar potential $\phi$ and a pseudo magnetic field $\vec{B}_\nu$ coupling to the
magnetic moment of the target. Using our approximations for one relativistic
neutrino species the relevant formulas are
\begin{align}
	\phi_{\nu,R}^D &\approx \phi_{\nu,R}^M \approx
	\sqrt{2} \, G_F \frac{\Delta_\nu}{\text{cm}^3} V_{11}  \approx -6.35 \times 10^{-38} \, \Delta_\nu \text{ eV} \;,
\end{align}
and for the pseudo magnetic fields coupling to the magnetic moment of a single neutron
\begin{align}
	|\vec{B}_{\nu,R}^D| &\approx |\vec{B}_{\nu,R}^M| \approx \sqrt{2} \, G_F \frac{1}{|\vec{\mu}_n|} \frac{|\Delta_\nu|}{\text{cm}^3}  A_{11} \approx 
	1.05 \times 10^{-30}  |\Delta_\nu| \text{ T} \;,
\end{align}
where we have used that the coupling of neutrinos to a neutron is flavor universal
and  $|\vec{\mu}_n| =  6.04 \times 10^{-14}$~MeV/T. Both values are tiny which makes it again
apparent how difficult it is to measure the CNB. For the scalar potential comes on top that
one could only measure a gradient of it, which could be induced dynamically due to refraction
but leads to another large suppression factor.

\subsubsection{Non-Relativistic Case}

The non-relativistic case is slightly more complicated as it also has a pseudo $B$-field
coupling to neutrinos and a spin-spin interaction, see eq.~\eqref{eq:VnuNR_phi_B}.
We do not discuss here the pseudo $B$-field acting
on neutrinos. This would be very difficult to measure despite it has a seemingly large value due to
the strong upper bound on the magnetic moment of a neutrino. 

Again we will just provide numbers here for one neutrino species coupling to a single neutron.
Using the approximations of the non-relativistic case we then find for Dirac neutrinos
\begin{align}
	\phi_{\nu,NR}^D &\approx   \frac{G_F}{\sqrt{2}} \frac{\Delta_\nu}{\text{cm}^3} V_{11} \approx -3.18 \times 10^{-38} \Delta_\nu \text{ eV}
	\;, \\
	|\vec{B}_{\nu,NR}^D| &\approx  \frac{G_F}{\sqrt{2} } \frac{1}{|\vec{\mu}_n|} \frac{|\Delta_\nu|}{\text{cm}^3} A_{11} 
	\approx 5.27 \times 10^{-31}  |\Delta_\nu| \text{ T} 
	\;, \\
	|I_{k,NR}^D| &\approx \frac{G_F}{\sqrt{2} } \frac{1}{|\vec{\mu}_k| |\vec{\mu}_n|} \frac{|\Delta_\nu|}{\text{cm}^3} A_{11} 
	\approx 5.27 \times 10^{-31}   \frac{|\Delta_\nu|}{|\vec{\mu}_k|} \text{ T} \;.
\end{align}	
In the expression for the spin-spin interaction we have kept the unknown $|\vec{\mu}_k|$ as a parameter.
If we only consider the coupling to the neutron magnetic moment, we can combine the spin-spin interaction and the
pseudo magnetic field. This would then look again like a pseudo magnetic field and it could double the size of
the original pseudo magnetic field.

For the Majorana case the formulas are even simpler as the pseudo magnetic field and the scalar potential
are zero and the effect on the neutron magnetic moment is coming completely from the spin-spin interaction
\begin{align}
	\phi_{\nu,NR}^M &= 0 \;, \\
	|\vec{B}_{\nu,NR}^M| &=  0 \;, \\
	|I_{k,NR}^M| &\approx \sqrt{2} \, G_F \frac{1}{|\vec{\mu}_k| |\vec{\mu}_n|} \frac{|\Delta_\nu|}{\text{cm}^3} A_{11} 
	\approx1.05 \times 10^{-30} \frac{|\Delta_\nu|}{|\vec{\mu}_k|} \text{ T} \;.
\end{align}
We have again kept the unknown $|\vec{\mu}_k|$ as an open parameter. Once we would apply all
the approximations from before $|\vec{\mu}_k|$ would be canceled and the spin-spin interaction
could also be written as an effective pseudo $B$-field acting on the neutron magnetic moment.

Again, like in the relativistic case, the scalar potentials and the magnetic fields are tiny and beyond what
can be currently measured in experiments as far as we know. So measuring the CNB in a laboratory on Earth
remains a formidable challenge.

\subsection{Dark Matter}

While this paper has a main focus on the CNB we want to briefly comment on DM as well.
There are actually already some papers on the subject, for instance, \cite{Du:2022ceh,Badurina:2025xwl},
which nevertheless take different approaches than ours. Here we will briefly discuss a DM induced
effective potential difference in the two interferometer arms which takes some strong inspiration
from the Dark Stodolsky effect~\cite{Rostagni:2023eic}. We also discuss a DM-matter potential
similar to the neutrino-matter potential. For a DM-neutrino coupling this could also lead to a 
Dark MSW effect.

A non-relativistic CNB forms a small component of DM in the universe and so we can easily recast the formulas
for this case for the DM scenario. As explicit example, we will focus here on one flavor of Majorana fermion DM, but Dirac
cases or even bosonic DM could have similar effects. For a complete list of potential operators, see~\cite{Rostagni:2023eic}.

We can get the potential for the considered DM case by rewriting eq.~\eqref{eq:Vnu_NR_M}
\begin{align}
	V_\text{DM} 
	&\approx  \sqrt{2} \, G_\text{DM}  \sum_{s_\text{DM} } n_\text{DM} \,  s_\text{DM}  \Biggl\{ V_\text{DM}  \frac{\vec{\mu}_\text{DM}  }{|\vec{\mu}_\text{DM} |} \cdot  \vec{\beta}_\text{DM}  -  s_T \, A_\text{DM}  \frac{\vec{\mu}_T }{|\vec{\mu}_T|} \cdot \frac{\vec{\mu}_\text{DM} }{|\vec{\mu}_\text{DM} |}  \Biggr\} \;.
\end{align}
where we include only vector and axial-vector type interactions and we set $|\vec{\beta}_T| = 0$. We also replaced the Fermi
constant, $G_F$, with an unknown constant $G_\text{DM}$ which, in principle, can be larger or smaller
than $G_F$.

The problem here is, that we do not know a lot about DM and most of the above parameters are
unknown or only weakly constrained. Take, for instance, the number density. We know from experiments
that the local DM energy density, $\rho_\text{DM}$, is about
$0.4$~GeV/cm$^3$~\cite{Catena:2009mf,Nesti:2013uwa,Sivertsson:2017rkp}. But since the DM
mass is unknown we do not know the number density. Here we want to focus on light DM with
somewhat large number densities and long wave-lengths which better justifies our approach.

Another issue is that without detailed model assumptions we cannot estimate the size of
$n_\text{DM}(\pm1)$ or even the average orientation of $\vec{\mu}_\text{DM}$, i.e., how well
it aligns with the DM momentum direction. In terms of a formula, we do not know the value of
\begin{align}
	\cos \theta_\text{DM} =  \frac{\vec{\mu}_\text{DM}  }{|\vec{\mu}_\text{DM} |} \cdot  \frac{ \vec{\beta}_\text{DM} }{| \vec{\beta}_\text{DM}|} \;.
\end{align}

Nevertheless, we provide here scaling formulas for the phase shifts 
on which more explicit models can be rather easily mapped
\begin{align}
	\Delta \Phi_{\text{DMM}}
		&\approx 1.48 \times 10^{-28} \frac{T}{1\text{ s}}  \frac{G_\text{DM}}{G_F} \frac{r_\text{DM}}{10^{-6}} \frac{\rho_\text{DM}}{0.4 \text{ GeV/cm}^3} \frac{1\text{ MeV}}{m_\text{DM}} \frac{a_\text{DM}}{0.5}   \frac{\cos \theta_\text{DM}}{0.5} \frac{F^V_\text{atom}}{10} 
	\;, \\
	\Delta \Phi_{\text{DST}} &\approx - 3.86 \times 10^{-19} \frac{T}{1\text{ s}}  \frac{G_\text{DM}}{G_F}  \frac{\rho_\text{DM}}{0.4 \text{ GeV/cm}^3} \frac{1\text{ MeV}}{m_\text{DM}}\frac{a_\text{DM}}{0.5}  \frac{F^A_\text{atom}}{10}  \;.
\end{align}
Here, we have introduced some new symbols. First of all, $r_\text{DM}$ is the analogue of
$r_\nu$. It is a dynamically generated gradient of the DM number densities. To make this happen, one would
need chiral interactions of DM with ordinary
matter, but then one can derive results in complete analogy to the neutrino case discussed
in \cite{Arvanitaki:2022oby,Huang:2024tog,Gruzinov:2024ciz,Kalia:2024xeq}. In fact, since DM
might have interactions stronger than the weak interaction at low energies, we use
a larger reference value for DM. 

Then we introduced $a_\text{DM}$ which is between $-1$ and $1$. It parametrizes the fraction
of DM which has $s_\text{DM} = \pm1$, to be precise,
\begin{align}
 	n_\text{DM} (+1)  -  n_\text{DM}(-1) = \frac{\rho_\text{DM} }{ m_\text{DM} }  a_\text{DM} \;.
\end{align}
Finally, we introduced the two factors $F^V_\text{atom}$ and $F^A_\text{atom}$. They 
parametrize the coupling of the vector and the axial vector to the atom. In the SM this
is determined by the structure of weak interactions. For instance, for the SM
$F^A_\text{atom}$ would be $(Z-N-E)/2$ which can be easily $\mathcal{O}(10)$.

\begin{figure}
	\centering
	\includegraphics[width=0.8\textwidth]{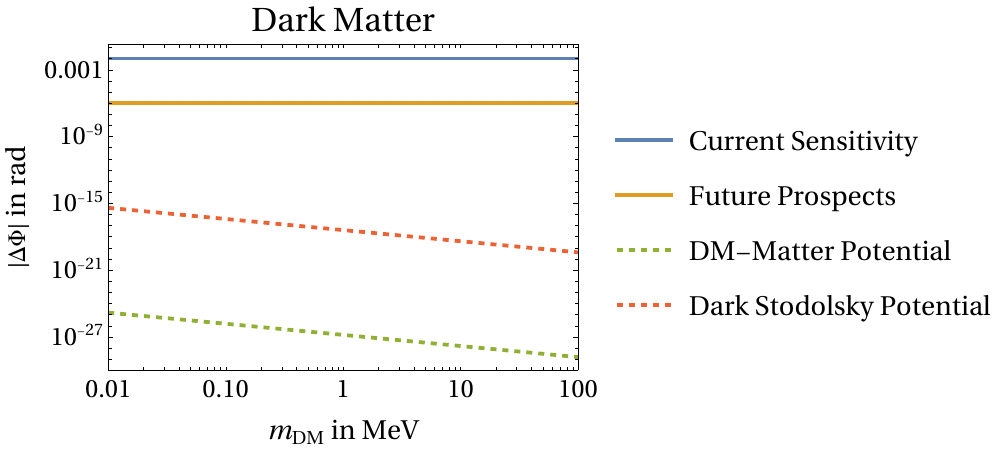} 
	\caption{\label{fig:DM}
		Dependence of $|\Delta \Phi|$ for the DM case with $T = 1$~s
		and $G_\text{DM} = 10 \, G_F$. Here we plot the effect of the DM-matter potential
		and the Dark Stodolsky potential depending on
		the DM mass. For more details, see main text. 
	}
\end{figure}

We have plotted the above numerical approximations in Fig.~\ref{fig:DM} as a function
of the DM mass $m_\text{DM}$. One thing to notice is that
the expectations go up for smaller DM masses. This is due to the fact that for smaller masses
the expected DM number density increases and hence the expected number of sources for the
potential go up as well. This is part of the reason why DM searches in interferometers
prefer lighter DM masses, for more details, see \cite{Arvanitaki:2016fyj,Du:2022ceh}.
Conventional optical interferometers actually already provide constraints on ultralight
(bosonic) DM, see, for instance, the recent result~\cite{LIGOScientific:2025ttj}.

Again the DM-matter potential (the analog to the neutrino-matter potential) is expected to a much
smaller phase shift since we again invoked a small local number density gradient induced by
diffraction effects near the surface of the Earth.

The Dark Stodolsky effect seems much more promising and is for our considered benchmark values
only a few orders of magnitude below expected sensitivities. Considering more extreme values
it might actually well be within future prospects of atom interferometers. This deserves a more detailed
study which also takes into account other experimental DM constraints. Such an enterprise is beyond the
scope of the current work which has the main focus on the CNB.

\subsection{CNB Scattering}

So far we did not discuss any scattering effects which are $\mathcal{O}(G_F^2)$.
Nevertheless, for completeness we want to comment on them briefly.
The scattering
cross section from the CNB from an atomic nucleus can be estimated as \cite{Formaggio:2012cpf,Domcke:2017aqj}
\begin{align}
	\sigma_{\text{CNB}} \approx \frac{G_F^2}{4 \, \pi \, \hbar^4 \, c^4} (Z-A)^2 E_\nu^2 \approx 10^{-63} \, (Z-A)^2 \text{ cm}^2\;.
\end{align}
If we multiply this with the CNB neutrino flux on Earth which is
$\mathcal{O}(10^{12} \text{ cm}^{-2} \text{ s}^{-1})$~\cite{Vitagliano:2019yzm} we end up with 
a scattering rate of $10^{-51}$~s$^{-1}$ for single neutrons. This
seems more challenging to be observed or separated from background 
even after including
coherence factors
compared to the other effects we discussed. So we do not discuss CNB scattering
in greater length here.

DM scattering might be more promising but this case was already discussed, for instance, in
\cite{Arvanitaki:2016fyj,Du:2022ceh} so we do not discuss it here further either.

\section{Summary and Conclusions}
\label{sec:Summary}

In this paper, we have discussed the prospects for detecting the CNB using matter interferometers.
In a previous paper~\cite{Domcke:2017aqj} a similar approach was taken for conventional
laser based interferometers. In that previous work the CNB was acting on the mirrors in the
interferometer but here there is a direct effect on the objects which are interfering.
In both cases the relevant physics is based on weak interactions and here we focused mainly
on the potentials which give an $\mathcal{O}(G_F)$ effect which is more promising than the
scattering $\mathcal{O}(G_F^2)$ effects as we mentioned.

The potential induced by the CNB on ordinary (fermionic) matter can be decomposed into two
components. First of all, there is the neutrino-matter potential which has a very well-known
implication, the MSW effect. The other part is maybe less well-known in the general physics community.
But for CNB studies it is known as the Stodolsky effect leading to a potential energy splitting between
the two spin-orientations of a fermion in the CNB bath. As we have shown here there is another way
to write this potential in terms of a scalar potential, pseudo magnetic fields and
a spin-spin interaction, see eqs.~\eqref{eq:VnuR_phi_B} and \eqref{eq:VnuNR_phi_B}, for relativistic
and non-relativistic neutrino species respectively. In any case, all effects only occur for an
asymmetric background. That can be a lepton asymmetry leading to different number densities
for neutrinos and anti-neutrinos or different number densities for different spin-orientations of the
neutrino background. In the SM these asymmetries are expected to be small but they can be
sizeable in SM extensions and the experimental upper bound is
$\mathcal{O}(10 \text{ cm}^{-3})$~\cite{Domcke:2025jiy}.

We then consider two different situations. First, we assume a matter interferometer where the
arms are spatially separated and due to diffraction effects near the surface of the Earth there can
be a gradient in the CNB number densities which drives the observable phase difference.
Unfortunately, the original potentials are quite small already and so is the expected gradient.
In this scenario depending on the neutrino mass scale and the nature of neutrinos we expect
phase differences of the order of $10^{-22}$ for neutrons and $10^{-20}$ for atoms. This is far too small to be seen for current
experiments with
sensitivities of the order of $10^{-2}$~rad. Even for the future when phase differences
as small as $10^{-6}$~rad or even $10^{-10}$~rad could be measured there is still
a considerable gap to fill.

In the second case we assume that the beam splitter acts as a polarization filter and thus it offers
the chance to measure the Stodolsky effect. This effect does not suffer from an additional suppression
by the gradient factor and is hence expected to be significantly larger. Again depending on the scenario
phase differences of the order of $10^{-14}$ for neutrons and $10^{-12}$ for atoms might be possible. This is still beyond the reach
of what we have seen in the literature but it might be possible to achieve using, for instance,
a significantly larger number of neutrons or atoms and a large interferometer time.
New physics effects which increase the lepton asymmetry in the early universe could also 
enhance the phase difference from the interaction potential.
It would be interesting to study a concrete model for that which is in agreement
with existing neutrino data but this goes beyond the scope of the current work.

Actually, there is a new physics scenario which is well motivated: Dark Matter. It is straight-forward
to apply our calculations to this case. For the Dark Stodolsky effect we find here plausible
phase differences of the order of $10^{-13}$ at $m_{\text{DM}} = 1$~MeV but there are much larger uncertainties due
to the unknown size of the couplings, DM asymmetries, etc. For $m_{\text{DM}} = 10$~keV an
induced phase difference
of the order of $10^{-10}$ seems actually quite possible. We leave a more detailed study for different
DM scenarios including other experimental DM constraints for future work as our main focus
here is on the CNB.

The detection of the CNB is known to be a very difficult issue and hence any new sensitive technology can
and should be considered for this purpose. 
That is what we did here for atom interferometers by providing some benchmark numbers for the required
sensitivity.
Unfortunately,
this idea (like all other proposals to detect the CNB) is unlikely to succeed with current technology.
Also, the next generation of detectors as far as we know will not achieve the necessary sensitivity.
But the numbers presented here can be seen as a
very appealing sensitivity goal for the future.
Measuring the CNB directly in a laboratory would allow us to gain insights into the properties of the 
universe in its very early
stage and we think it is important to explore possible ideas to achieve this goal. Furthermore, our formalism
and approach can easily be generalized to a wide class of DM models where the prospects are more
promising and relevant for atom interferometers rather soon.

\section*{Acknowledgements}

Chrisna Setyo Nugroho is supported by the National Science and Technology Council of Taiwan under Grant No.\ NSTC 114-2112-M-003-009 (114B0263).
Martin Spinrath is supported by the National Science and Technology Council (NSTC) of Taiwan under
Grant No.\ NSTC 114-2112-M-007-030.

\end{document}

%% file: figs/Setup.tikz
\tikzset{every picture/.style={line width=0.75pt}}    

\begin{tikzpicture}[x=0.75pt,y=0.75pt,yscale=-1,xscale=1]

\draw    (240,220) -- (240,172) ;
\draw [shift={(240,170)}, rotate = 90] [color={rgb, 255:red, 0; green, 0; blue, 0 }  ][line width=0.75]    (10.93,-3.29) .. controls (6.95,-1.4) and (3.31,-0.3) .. (0,0) .. controls (3.31,0.3) and (6.95,1.4) .. (10.93,3.29)   ;

\draw    (240,120) -- (288,120) ;
\draw [shift={(290,120)}, rotate = 180] [color={rgb, 255:red, 0; green, 0; blue, 0 }  ][line width=0.75]    (10.93,-3.29) .. controls (6.95,-1.4) and (3.31,-0.3) .. (0,0) .. controls (3.31,0.3) and (6.95,1.4) .. (10.93,3.29)   ;

\draw    (340,220) -- (340,172) ;
\draw [shift={(340,170)}, rotate = 90] [color={rgb, 255:red, 0; green, 0; blue, 0 }  ][line width=0.75]    (10.93,-3.29) .. controls (6.95,-1.4) and (3.31,-0.3) .. (0,0) .. controls (3.31,0.3) and (6.95,1.4) .. (10.93,3.29)   ;

\draw    (240,170) -- (240,120) ; 
\draw    (340,170) -- (340,120) ; 

\draw    (240,220) -- (288,220) ;
\draw [shift={(290,220)}, rotate = 180] [color={rgb, 255:red, 0; green, 0; blue, 0 }  ][line width=0.75]    (10.93,-3.29) .. controls (6.95,-1.4) and (3.31,-0.3) .. (0,0) .. controls (3.31,0.3) and (6.95,1.4) .. (10.93,3.29)   ;

\draw    (170,220) -- (208,220) ;
\draw [shift={(210,220)}, rotate = 180] [color={rgb, 255:red, 0; green, 0; blue, 0 }  ][line width=0.75]    (10.93,-3.29) .. controls (6.95,-1.4) and (3.31,-0.3) .. (0,0) .. controls (3.31,0.3) and (6.95,1.4) .. (10.93,3.29)   ;

\draw    (340,120) -- (378,120) ;
\draw [shift={(380,120)}, rotate = 180] [color={rgb, 255:red, 0; green, 0; blue, 0 }  ][line width=0.75]    (10.93,-3.29) .. controls (6.95,-1.4) and (3.31,-0.3) .. (0,0) .. controls (3.31,0.3) and (6.95,1.4) .. (10.93,3.29)   ;

\draw    (290,120) -- (340,120) ;
\draw    (290,220) -- (340,220) ;
\draw    (210,220) -- (240,220) ;
\draw    (380,120) -- (410,120) ;

\draw    (211.06,189) -- (271,249)(209,191) -- (269,251) ;
\draw    (211.06,89) -- (371,249)(209,91) -- (369,251) ;
\draw    (311.06,89) -- (371,149)(309,91) -- (369,151) ;

\draw (229,223) node [anchor=north west][inner sep=0.75pt]   [align=left] {$A$};
\draw (326,223) node [anchor=north west][inner sep=0.75pt]   [align=left] {$B_2$};
\draw (240,103) node [anchor=north west][inner sep=0.75pt]   [align=left] {$B_1$};
\draw (341,103) node [anchor=north west][inner sep=0.75pt]   [align=left] {$C$};

\draw    (115.28,208.5) -- (170.28,208.5) -- (170.28,232.5) -- (115.28,232.5) -- cycle  ;
\draw (142.78,220.5) node   [align=left] {Source};

\draw    (410.36,108.5) -- (474.36,108.5) -- (474.36,132.5) -- (410.36,132.5) -- cycle  ;
\draw (442.36,120.5) node   [align=left] {Detector};

\draw (257,252) node [anchor=north west][inner sep=0.75pt]   [align=left] {$t = 0$};
\draw (355,252) node [anchor=north west][inner sep=0.75pt]   [align=left] {$t = T$};
\draw (375,145) node [anchor=north west][inner sep=0.75pt]   [align=left] {$t = 2 \, T$};
\draw (199,171) node [anchor=north west][inner sep=0.75pt]    {$\pi/2$};
\draw (300,70) node [anchor=north west][inner sep=0.75pt]    {$\pi/2$};
\draw (199,76) node [anchor=north west][inner sep=0.75pt]    {$\pi$};

\end{tikzpicture}